\documentclass[
]{ceurart}

\hypersetup{
    colorlinks=false,
    pdfborder={0 0 1},
    citebordercolor={1 0 0},
    linkbordercolor={0 1 0},
    urlbordercolor={0 0 1}
}

\usepackage{listings}
\usepackage{xurl}
\begin{document}


\copyrightyear{2026}
\copyrightclause{Copyright for this paper by its authors.
  This author version is made available under the Creative Commons
  Attribution 4.0 International License (CC BY 4.0).}

\conference{Preprint of a paper accepted at the 1st Workshop on Extraction from
Triplet Text-Table-Knowledge Graph and associated Challenge (TRIPLET),
co-located with ESWC 2026, May 10--14, 2026, Dubrovnik, Croatia}

\title{Towards Foundation Models for Relational Databases with Language Models and Graph Neural Networks}

\author[1]{Jingcheng Wu}[%
email=jingcheng.wu@ki.uni-stuttgart.de,
]
\author[1]{Ratan Bahadur Thapa}[%
email=ratan.thapa@ki.uni-stuttgart.de,
]
\author[1]{Mojtaba Nayyeri}[%
email=mojtaba.nayyeri@ki.uni-stuttgart.de,
]
\author[1]{Lucas Etteldorf}[%
email=st191540@stud.uni-stuttgart.de,
]
\author[1]{Max Finkenbeiner}[%
email=st191433@stud.uni-stuttgart.de,
]
\author[1]{Fabian Leeske}[%
email=st169266@stud.uni-stuttgart.de,
]
\author[1,2]{Steffen Staab}[%
email=steffen.staab@ki.uni-stuttgart.de,
]

\address[1]{University of Stuttgart, Stuttgart, Germany}
\address[2]{Web and Internet Science Research Group, University of Southampton, Southampton, United Kingdom}

\begin{abstract}
  Relational databases store much of the world's structured information, and they are essential for driving complex predictive applications. However, deep learning progress on relational data remains limited, as conventional approaches flatten databases into single tables via manual feature engineering, discarding relational context. Relational deep learning (RDL) addresses this by modeling databases as relational entity graphs (REGs) for graph neural networks (GNNs), but remains task- and database-specific. To combine the strengths of both paradigms, we propose a hybrid architecture combining a fine-tuned BART encoder to capture intra-row semantics with a GraphSAGE-based GNN over REGs to inject relational context. Experiments on RelBench show that the GNN substantially enriches BART's row embeddings, achieving a ROC-AUC of 67.40 on the \texttt{driver-dnf} task from the \textit{rel-f1} dataset. This performance is competitive with supervised baselines such as LightGBM (68.86) and narrows the gap to RDL (72.62) to within 5.22 points, though a substantial gap remains to state-of-the-art foundation models such as KumoRFM (82.63). These results suggest that lightweight hybrid LM-GNN architectures offer a promising and resource-efficient path towards foundation models for relational databases.
\end{abstract}


\maketitle

\section{Introduction}
\label{sec:introduction}

Relational databases store much of the world's structured data in multiple tables, with rows in different tables connected using primary and foreign keys \cite{fey2024position,dwivedi2025relational}.
They underpin large-scale information systems in domains such as e-commerce, banking, and healthcare. Many real-world predictive tasks depend on such relational data, including forecasting future product sales, predicting user churn, or estimating the risk of discharging a patient. To perform such tasks effectively, it is essential to leverage both the data in the tables and the relationships encoded in the database schema.

Although the success of deep learning methods on unstructured data makes its application to relational databases highly desirable, these methods have inherently struggled with the complexity of relational structures \cite{fey2024position}. Since existing tabular models cannot directly learn across multiple interconnected tables, the standard practice relies on manual feature engineering. Relational databases are typically flattened into a single table through joins and aggregations before applying conventional machine learning methods. This flattening procedure requires domain expertise, is slow and error-prone, and severely degrades the relational context provided by primary-foreign key links.

To address these limitations, Fey et al.\ \cite{fey2024position} introduced relational deep learning (RDL), which models databases as relational entity graphs (REGs) and applies graph neural networks (GNNs). Although RDL improves upon feature engineering across most RelBench tasks, its models are trained separately per database and task, and do not learn transferable representations. While this limitation has recently motivated the first relational foundation models \cite{wang2025griffin,fey2025kumo}, these architectures are highly complex and computationally intensive, posing significant barriers to widespread adoption.

Seeking a more lightweight and accessible alternative, we draw inspiration from the hybrid LM-GNN concept initially proposed by Vogel et al.\ \cite{vogel2023towards}. As their framework focused exclusively on generation-based data preparation tasks, such as missing value imputation, it cannot be directly applied to general-purpose predictive transfer. To bridge this gap, we propose an extended architecture combining two core components: a fine-tuned BART encoder to extract row-level semantics, and a GraphSAGE-based GNN over REGs to inject relational context. To assess whether this hybrid approach can serve as a foundation towards general-purpose relational foundation models, we evaluate our architecture on the \texttt{driver-dnf} node classification task from the \textit{rel-f1} dataset in RelBench.

Our main contributions are as follows:
\begin{enumerate}
    \item We propose a novel LM-GNN hybrid architecture that seamlessly integrates a pre-trained language model (BART) with a GraphSAGE-based GNN. The language model captures deep intra-row semantics, while the GNN enriches these embeddings with structural relational context. Coupled with self-supervised objectives, namely schema-aware token masking for the LM and masked feature reconstruction for the GNN, this framework generates contextualized representations easily adaptable to downstream tasks with minimal computational overhead.
    \item We provide a systematic empirical evaluation of our architecture on the RelBench benchmark. Our model achieves a highly competitive ROC-AUC score of 67.40 on the \texttt{driver-dnf} task from the \textit{rel-f1} dataset. Furthermore, our extensive ablation study demonstrates that explicitly incorporating relational context via message passing yields substantial ROC-AUC improvements compared to relying solely on isolated BART or random embeddings, highlighting the critical necessity of structural message passing.
    \item We outline a strategic roadmap for future relational foundation models based on an in-depth analysis of our framework's strengths and limitations. Specifically, we identify key research directions, including the joint training of the LM-GNN pipeline and architectural adaptations required to scale across diverse relational databases.
\end{enumerate}

\section{Background}
\label{sec:background}

We adopt the formalism of Fey et al.\ \cite{fey2024position}. A relational database $(\mathcal{T}, \mathcal{L})$ consists of tables $\mathcal{T} = \{T_1, \dots, T_n\}$ and links $\mathcal{L} \subseteq \mathcal{T} \times \mathcal{T}$, where a link $(T_{\text{fkey}}, T_{\text{pkey}}) \in \mathcal{L}$ exists when a foreign key column of $T_{\text{fkey}}$ references the primary key of $T_{\text{pkey}}$. Each table $T = \{v_1,\dots,v_{n_T}\}$ contains entities (rows), each with a primary key $p_v$, foreign keys $K_v$, attributes $x_v = \{x_v^1,\dots,x^{d_T}_v\}$, and an optional timestamp $t_v$. All entities in the same table share the same columns, though individual values may differ or be missing.

\paragraph{Schema Graph and Relational Entity Graph.}
The \textit{schema graph} $(\mathcal{T}, \mathcal{R})$ captures the table-level structure, with tables as nodes and bidirectional edges $\mathcal{R} = \mathcal{L} \cup \mathcal{L}^{-1}$ (where $\mathcal{L}^{-1}$ contains the inverse of each link). The \textit{relational entity graph} (REG) is a heterogeneous graph $G = (\mathcal{V}, \mathcal{E}, \phi, \psi)$ suitable for GNN processing. Its node set $\mathcal{V} = \bigcup_{T \in \mathcal{T}} T$ contains all database rows, and its edge set $\mathcal{E}$ connects entity pairs linked by primary-foreign key relationships. Type mapping functions $\phi: \mathcal{V} \to \mathcal{T}$ and $\psi: \mathcal{E} \to \mathcal{R}$ assign each node and edge to its corresponding schema graph element, enabling type-specific handling by GNNs. In our architecture, we use node-type-specific linear layers (Section~\ref{subsec:architecture}). Each node $v \in \mathcal{V}$ carries an embedding $h_v \in \mathbb{R}^d$. Initial embeddings are obtained from our fine-tuned BART encoder (Section~\ref{subsubsec:fine-tuning-bart}) and enriched via the GNN (Section~\ref{subsubsec:training-gnn}).

\section{Related Work}
\label{sec:related-work}

\paragraph{Foundation Models and Tabular Learning.}
Foundation models are large models pre-trained on broad data with self-supervised objectives and adaptable to diverse downstream tasks via fine-tuning \cite{Bommasani2021FoundationModels}. They have profoundly transformed NLP and vision. GPT-3 demonstrated that scaling enables in-context learning and few-shot generalization \cite{brown2020language}, while models such as CLIP and its variants \cite{radford2021clip,DBLP:conf/aaai/ZhouHMSZWNS26} extend these principles to multimodal settings. Their effectiveness stems from self-supervised representation learning \cite{ericsson2022self} combined with transfer learning across datasets and tasks \cite{zhuang2020comprehensive}.

Despite these successes in unstructured domains, deep learning has long underperformed tree-based methods such as XGBoost \cite{chen2016xgboost} due to challenges including small dataset sizes, absence of spatial or sequential structure, and heterogeneous feature types mixing numerical and categorical attributes \cite{borisov2022deep,grinsztajn2022tree}. Recent transformer-based methods, including TabTransformer \cite{huang2020tabtransformer}, FT-Transformer \cite{gorishniy2021revisiting}, and SAINT \cite{somepalli2021saint}, have narrowed this gap via attention mechanisms that capture complex feature interactions. More recently, TabPFN \cite{hollmann2025accurate} emerged as a tabular foundation model that outperforms prior methods on datasets up to 10,000 samples. However, all these advances remain limited to single tables and cannot exploit the relational structure of multi-table databases.

\paragraph{Language Models for Structured Data.}
A parallel line of work adapts pre-trained language models to structured inputs by linearizing rows and schema information. At the core of many such approaches are foundational models like BART \cite{lewis2019bart}. As a sequence-to-sequence model pre-trained as a denoising autoencoder, BART produces rich contextual embeddings by reconstructing corrupted input sequences. For single tables, TaBERT \cite{yin2020tabert} jointly models tabular inputs and textual inputs, while TURL \cite{deng2020turl} learns contextualized table representations. Extending to relational databases, RPT \cite{tang2020rpt} applies sequence-to-sequence models to data preparation tasks such as missing value imputation. While these works demonstrate the feasibility of language models for tabular data, they remain fundamentally limited to isolated tables or data preparation procedures rather than general-purpose predictive transfer.

\paragraph{Graph Neural Networks and Relational Deep Learning.}
GNNs extend deep learning to graph-structured data, where message-passing architectures iteratively update node embeddings by aggregating neighborhood information \cite{dwivedi2025relational,kipf2016semi,DBLP:conf/emnlp/DingWW0XT24}. GraphSAGE \cite{hamilton2017inductive} introduced inductive learning via neighborhood sampling, enabling generalization to unseen nodes. This property is crucial for the development of foundation models. Self-supervised objectives such as GraphMAE \cite{hou2022graphmae}, a masked graph autoencoder, further improve generalization and inspired our pre-training procedure.

RDL \cite{fey2024position} applies these ideas to relational databases by representing them as temporal, heterogeneous REGs and training GNNs end-to-end. It automates the construction of training labels from the database itself, encodes rows into initial node features, applies message-passing GNNs, and trains task-specific prediction heads. While RDL outperforms manual feature engineering with reduced development time, it remains database-specific and task-specific. These models are trained separately for each setting and lack the capacity to transfer across domains \cite{dwivedi2025relational}.

\paragraph{Foundation Models for Relational Databases.}
Building on RDL, recent efforts mark the first steps towards relational foundation models. Griffin \cite{wang2025griffin} follows a graph-centric approach utilizing unified input encoders for diverse data types, alongside standardized task decoders. Its advanced architecture incorporates cross-attention for node feature aggregation alongside relation-aware message passing. After pre-training on diverse single-table datasets and joint supervised fine-tuning on selected RDB subsets (the Griffin-RDB-SFT variant in \cite{wang2025griffin}), followed by task-specific fine-tuning, Griffin achieves an average ROC-AUC of 75.00 on classification tasks. For the zero-shot relational baseline in Section~\ref{subsec:results}, we report the Griffin-pretrained variant (pre-trained exclusively on single-table datasets, as defined in~\cite{wang2025griffin}).

KumoRFM \cite{fey2025kumo}, a proprietary commercial model, encodes database rows via a shared encoder tailored to semantic column types. It then applies a relational graph transformer to the resulting graph representation. This design enables in-context learning, allowing KumoRFM to generalize to previously unseen databases and tasks. Without prior training on test datasets, it achieves an average classification ROC-AUC of 76.71 and reaches 81.14 after fine-tuning.

Our work builds on the earlier vision of Vogel et al.\ \cite{vogel2023towards}, who proposed combining an LM encoder for row embeddings with a GNN for relational context. Their prototype demonstrated promise on data engineering tasks such as missing value imputation but did not address predictive tasks or broader generalization. We extend this vision by evaluating the hybrid architecture in a predictive setting on RelBench, positioning it as a lightweight, accessible alternative to the complex architectures of Griffin and KumoRFM.

\section{Methodology}
\label{sec:methodology}

We extend the hybrid LM-GNN architecture of Vogel et al.\ \cite{vogel2023towards} by integrating RDL principles \cite{fey2024position} and introducing self-supervised pre-training objectives. We represent relational databases as REGs, with initial node features obtained from a fine-tuned BART encoder \cite{lewis2019bart} and subsequently enriched via message passing within a heterogeneous GNN. This two-stage pipeline first captures row-level semantics via BART fine-tuning, and then injects relational dependencies through the GNN's message passing, optimized via self-supervised pre-training.

\subsection{Architecture}
\label{subsec:architecture}
\begin{figure}[ht]
    \centering
    \includegraphics[width=1.0\textwidth]{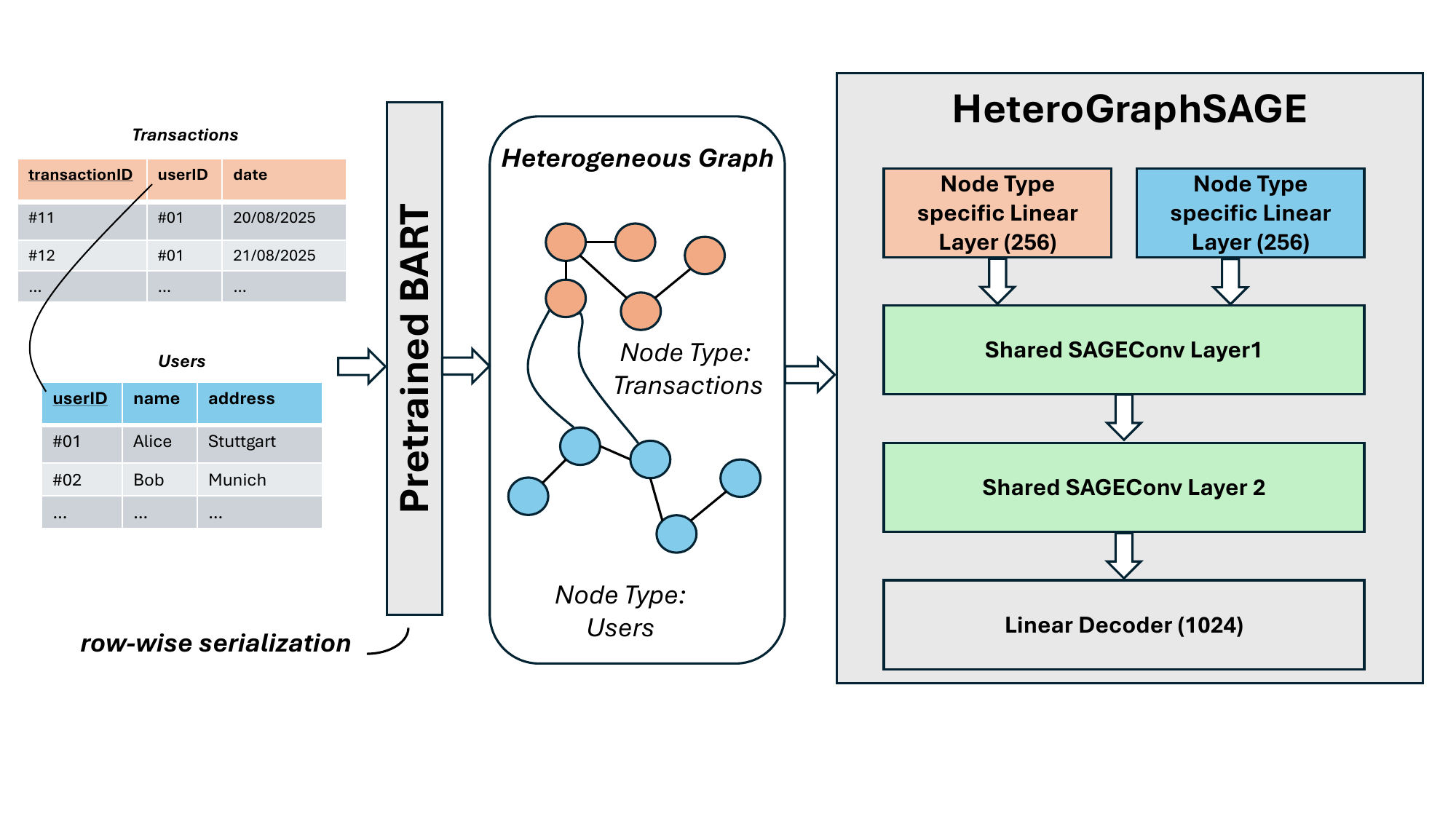}
    \caption{Overview of the hybrid architecture. A fine-tuned BART encoder generates row-level embeddings from linearized database rows, which serve as initial node features in the relational entity graph (REG). Node-type-specific linear layers project the 1024-dimensional BART embeddings to the 256-dimensional hidden space. Two shared SAGEConv layers then perform message passing across all edge types, and a linear decoder maps the enriched embeddings back to 1024 dimensions for reconstruction loss computation.}
    \label{fig:approach-setup}
\end{figure}

As illustrated in Fig.~\ref{fig:approach-setup}, our framework consists of two components: (1)~\textbf{a fine-tuned BART encoder} that linearizes database rows into schema-aware strings and produces row-level embeddings capturing intra-row semantics, and (2)~\textbf{a GraphSAGE-based GNN} operating on REGs that enriches these embeddings with relational context via message passing.

We choose BART for its denoising pre-training that yields rich contextual embeddings and its demonstrated effectiveness on relational tasks \cite{vogel2023towards}. For the GNN, we adopt GraphSAGE \cite{hamilton2017inductive} for its inductive generalization to unseen nodes, scalable mini-batch training, and strong benchmark performance \cite{dwivedi2022}. To handle heterogeneous graph data, we employ a HeteroGraphSAGE model via PyTorch Geometric \cite{DBLP:journals/corr/abs-1903-02428} with node-type-specific linear layers and \textit{shared} \texttt{SAGEConv} layers across all edge types to improve generalization to unseen schemas. The GNN output dimensionality is set to 1024, matching the BART embeddings for reconstruction loss computation.

\subsection{Data Preparation}
\label{subsec:data-preparation}

We use the RelBench benchmark \cite{robinson2024relbench}, selecting 6 of its 7 databases (\textit{rel-trial, rel-avito, rel-hm, rel-amazon, rel-event, rel-stack}) for pre-training and reserving \textit{rel-f1} for downstream evaluation to ensure an unbiased generalization test.

\paragraph{Preprocessing for BART.}
We randomly sample 100,000 rows evenly distributed across the 6 databases and their tables. Each row is linearized into a schema-aware sequence:
\begin{center}
    \text{<table> Table Name <attr> Attr1 Name <value> Attr1 Value} \dots
\end{center}
The special tokens \texttt{<table>}, \texttt{<attr>}, and \texttt{<value>} are added to BART's tokenizer vocabulary as new tokens, and the model's token embeddings are resized accordingly. Missing and NaN values are replaced with empty strings. This format encodes table names, attribute names, and values, enabling BART to capture schema-data dependencies. The resulting rows are split into a 70/10/20 ratio for training, validation, and testing sets.

\paragraph{Preprocessing for the GNN.}
Each database is transformed into a heterogeneous REG, with tables as distinct node types and primary-foreign key relationships as edges. Rows are encoded using the fine-tuned BART model, yielding 1024-dimensional node embeddings. For computational feasibility, larger tables are subsampled to limit graph size, producing graphs of approximately 4 to 6 million nodes each and totaling around 28 million nodes and 40 million edges across all six databases (Table~\ref{tab:relbench-datasets}).

\begin{table}[htbp]
\centering
\caption{Statistics of the six RelBench databases used for GNN pre-training. Node and edge counts reflect the truncated versions used in our experiments (see Section~\ref{subsec:data-preparation}).}
\label{tab:relbench-datasets}
\begin{tabular}{lcc}
\toprule
\textbf{Dataset} & \textbf{Nodes} & \textbf{Edges} \\
\midrule
rel-amazon    & 4,589,766 & 7,000,000 \\
rel-avito     & 5,142,166 & 9,488,416 \\
rel-event     & 6,001,744 & 6,020,839 \\
rel-hm        & 4,047,698 & 7,000,000 \\
rel-trial     & 4,921,015 & 4,671,285 \\
rel-stack     & 4,023,229 & 5,870,601 \\
\midrule
Total &\textbf{28,725,618}&\textbf{40,051,141}\\
\bottomrule
\end{tabular}
\end{table}

\subsection{Training Pipeline}
\label{subsec:training-pipeline}

\subsubsection{Stage 1: Fine-tuning BART}
\label{subsubsec:fine-tuning-bart}
We fine-tune a pre-trained BART model\footnote{\href{https://huggingface.co/facebook/bart-large}{We use the \texttt{facebook/bart-large} checkpoint from Hugging Face.}} \cite{lewis2019bart} on the linearized rows, which are tokenized and truncated to a maximum sequence length of 1024 tokens. Instead of standard token-level masking, we apply a schema-aware masking strategy at the level of semantic units. Table names are masked with probability 0.30, attribute names with 0.20, and cell values with 0.40. At least one element is masked per row, and empty values are skipped. The model is trained to reconstruct the original sequence using cross-entropy loss.

Fine-tuning proceeds for 50 epochs with the AdamW optimization, a cosine learning rate schedule with warmup, and an effective batch size of 128. Masking is applied dynamically per epoch on the training set. Validation and test sets use static masks. The resulting encoder produces 1024-dimensional row embeddings via mean-pooling of the last hidden states of the tokens, which serve as initial node features for the GNN stage.

\subsubsection{Stage 2: Training the GNN}
\label{subsubsec:training-gnn}
The GNN is trained with a \textit{masked value reconstruction} objective. For each node, random feature dimensions are masked and the loss is computed only on these dimensions, forcing the model to infer missing information via message passing from neighbors. Unlike standard reconstruction objectives that consider the entire embedding vector, restricting the loss to masked dimensions prevents the model from trivially copying observed values and explicitly encourages it to leverage relational context. We adopt a weighted combination of scaled cosine error \cite{hou2022graphmae} and MSE:
\begin{equation}
    \mathcal{L} = \alpha \, \mathcal{L}_{\text{cos}} + (1-\alpha) \, \mathcal{L}_{\text{mse}},
\end{equation}
where, for masked dimensions $\mathcal{M}$ of each node $i$ with true features $\mathbf{x}_i$ and reconstruction $\hat{\mathbf{x}}_i$:
\begin{equation}
    \mathcal{L}_{\text{cos}} = \frac{1}{N} \sum_{i=1}^{N}
    \left( 1 - \frac{\hat{\mathbf{x}}_i^{\mathcal{M}} \cdot \mathbf{x}_i^{\mathcal{M}}}
    {\|\hat{\mathbf{x}}_i^{\mathcal{M}}\| \, \|\mathbf{x}_i^{\mathcal{M}}\| + \epsilon} \right)^{\gamma}, \quad
    \mathcal{L}_{\text{mse}} = \frac{1}{N} \sum_{i=1}^{N}
    \|\hat{\mathbf{x}}_i^{\mathcal{M}} - \mathbf{x}_i^{\mathcal{M}}\|_2^2.
\end{equation}
Here $\epsilon = 10^{-6}$ ensures numerical stability, $\gamma = 2$ emphasizes larger cosine errors following Hou et al.\ \cite{hou2022graphmae}, and $N$ is the batch size.

Training utilizes the \texttt{NeighborLoader} utility from PyTorch Geometric \cite{DBLP:journals/corr/abs-1903-02428} with a neighborhood sampling strategy of (20, 10) (up to 20 first-hop and 10 second-hop neighbors per node) with mean aggregation and self-loops. To prevent catastrophic forgetting across databases, we interleave batches from all six sources during training, inspired by replay strategies from graph continual learning \cite{zhang2024}.

\begin{table}[ht]
\centering
\caption{Effect of masking probability on reconstruction MSE loss (test set). The best result is shown in bold.}
\label{tab:masking}
\begin{tabular}{lc}
\toprule
\textbf{Masking Prob.} & \textbf{MSE Loss} \\
\midrule
\textbf{0.15} & $\boldsymbol{8.90 \times 10^{-5}}$ \\
0.25 & $9.80 \times 10^{-5}$ \\
0.50 & $1.19 \times 10^{-4}$\\
1.00 & $1.73 \times 10^{-4}$\\
\bottomrule
\end{tabular}
\end{table}

\begin{figure}[ht]
    \centering
    \includegraphics[width=1.0\linewidth]{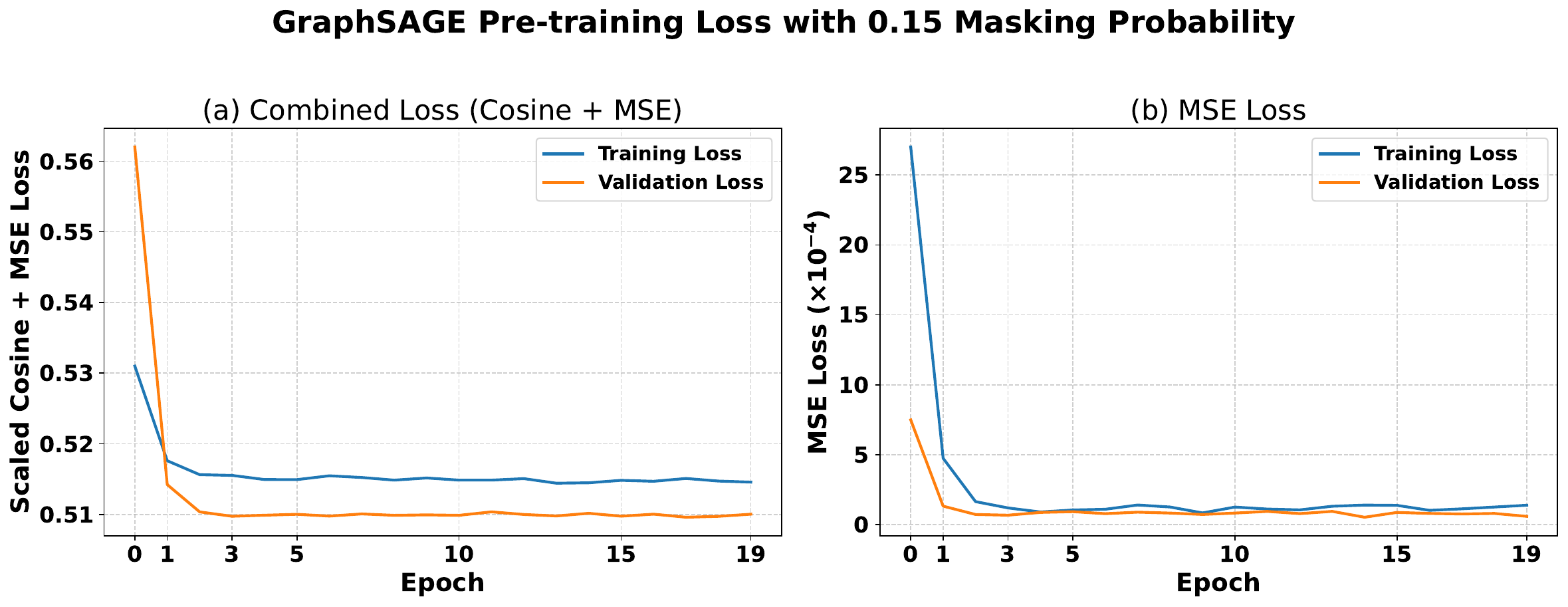}
    \caption{Loss curves during GNN pre-training with a masking probability of 0.15. (a) Combined scaled cosine and MSE loss for the training and validation sets. (b) Training MSE loss. The model largely converges by approximately epoch 5.}
    \label{fig:loss-pre-training}
\end{figure}

We select a masking probability of 0.15 based on a hyperparameter search (Table~\ref{tab:masking}). Lower probabilities yield more stable training and consistently better reconstruction, while higher probabilities tend to oversmooth predictions and diminish the model's ability to recover fine-grained feature information. The final model uses 256 hidden channels, two shared SAGEConv layers, a batch size 16,384, the Adam optimizer (lr = $1.0 \times 10^{-4}$), and $\alpha = 0.7$. We set $\alpha = 0.7$ to prioritize cosine similarity, which captures directional alignment between embeddings, while retaining MSE as a complementary magnitude-sensitive term. This weighting was selected based on preliminary experiments on the validation set.  Training largely converges by epoch 5 (Fig.~\ref{fig:loss-pre-training}), and we report a final test MSE of $8.90 \times 10^{-5}$ after the full 20-epoch schedule.

\subsection{Downstream Adaptation}
\label{subsec:downstream-adaptation}

For downstream tasks, new node-type-specific linear layers for the target schema and a lightweight prediction head are always trained from scratch. The shared SAGEConv layers transferred from pre-training are treated in one of two ways: (1)~\textbf{Frozen}: the shared SAGEConv weights are fixed, enabling rapid adaptation with minimal overhead; (2)~\textbf{Fine-tuned}: the shared SAGEConv weights are updated jointly with the prediction head, allowing the model to specialize its relational reasoning to the new database schema. Both approaches rely on the exact same pipeline: constructing the REG, obtaining BART embeddings, and propagating them through the GNN.

\section{Experiments}
\label{sec:experiments}

To evaluate whether our hybrid LM-GNN architecture can generalize to unseen relational databases, we conduct experiments on a held-out dataset from RelBench. We first describe the experimental setup, including the dataset, task, baselines, and training procedure (Section~\ref{subsec:experimental-setup}). We then present and analyze the main results in comparison to supervised methods and relational foundation models (Section~\ref{subsec:results}). Finally, we perform an ablation study to isolate the contributions of the BART encoder and the GNN component (Section~\ref{subsec:ablation-study}).

\subsection{Experimental Setup}
\label{subsec:experimental-setup}

\paragraph{Dataset and Task.}
We evaluate our framework on the held-out \textit{rel-f1} database, which comprises 9 relational tables (including \texttt{drivers}, \texttt{results}, and \texttt{races}) with 97,606 rows. Its small size and distinct domain make it suitable for testing generalization. We focus on the \texttt{driver-dnf} node classification task, predicting, for a given driver identifier and timestamp, whether the driver will fail to finish at least one race within the subsequent one-month window. We report ROC-AUC as the primary metric following RelBench conventions, alongside precision, accuracy, and F1-score in our ablation study.

\begin{figure}[ht]
    \centering
    \includegraphics[width=1.0\textwidth]{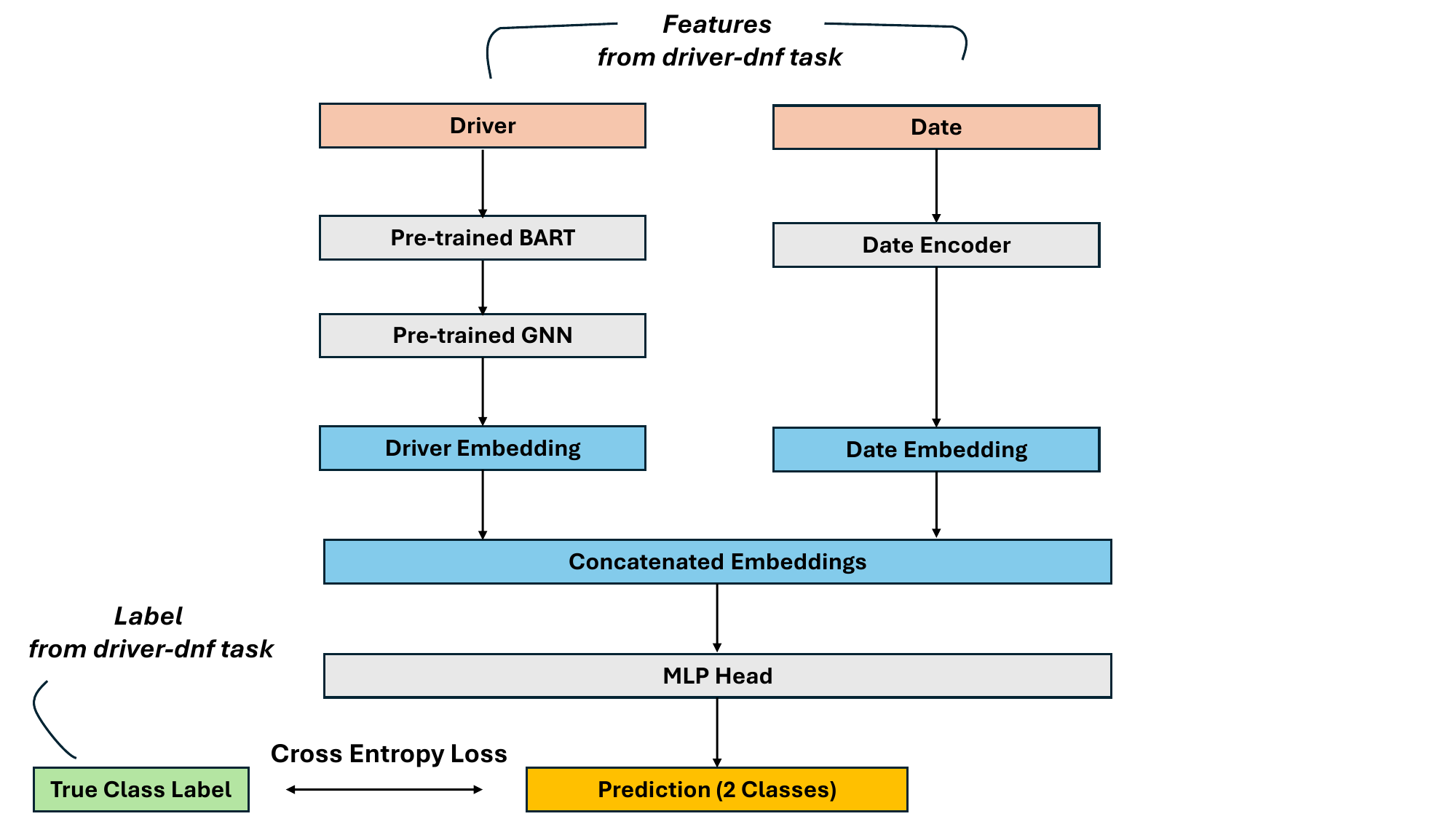}
    \caption{Downstream adaptation pipeline for the \texttt{driver-dnf} classification task. The driver identifier is encoded by the pre-trained BART encoder and enriched via the pre-trained GNN, while the task date is encoded by a separate date encoder. The resulting embeddings are concatenated and passed through an MLP head trained with cross-entropy loss. Color coding: light red denotes task input features, grey represents learnable or pre-trained network components, blue indicates intermediate embeddings, orange highlights model predictions, and green shows ground-truth labels.}
    \label{fig:experiment-setup}
\end{figure}

\paragraph{Baselines.}
We benchmark our framework against several supervised methods: LightGBM \cite{ke2017lightgbm}, manual feature engineering by data scientists \cite{robinson2024relbench}, and RDL \cite{fey2024position}, which serves as a supervised, per-database GNN baseline. Additionally, we evaluate against recent relational foundation models, including the Griffin-pretrained variant from \cite{wang2025griffin}(pre-trained exclusively on single-table datasets, evaluated without RDB-specific fine-tuning), and KumoRFM \cite{fey2025kumo}, a commercial model evaluated in both in-context and fine-tuned settings. Baseline scores are sourced from their respective original papers, with the exception of the Griffin-pretrained variant, which we evaluated locally using the official open-source implementation. Given that relational foundation models are only beginning to emerge with scarce public implementations, this evaluation provides a representative snapshot of the current landscape rather than a strictly controlled comparison.

\paragraph{Training.}
Following Section~\ref{subsec:downstream-adaptation}, all rows are encoded via the fine-tuned BART model to construct the initial REG. We transfer the shared SAGEConv layers from the pre-training phase, while freshly initializing new node-type-specific linear layers for the unseen target schema. In the frozen setting, only these new type-specific layers, the date encoder, and the MLP head are updated; in the fine-tuned setting, the transferred SAGEConv layers are additionally updated. As illustrated in Fig.~\ref{fig:experiment-setup}, the target driver node embeddings are enriched through the GNN. In parallel, the task date is processed by a 32-dimensional encoder consisting of two linear layers with ReLU activations. These two representations are then concatenated and fed into an MLP classification head, which comprises three linear layers equipped with ReLU and dropout. The entire downstream architecture is optimized using cross-entropy loss. For this predictive task, we directly utilize the 256-dimensional GNN output rather than projecting it back to 1024 dimensions as was required during pre-training.

\subsection{Results and Analysis}
\label{subsec:results}

Table~\ref{tab:results} reports the test ROC-AUC scores on the \texttt{driver-dnf} task.

\begin{table}[ht]
    \centering
    \caption{Test ROC-AUC on the driver-dnf task (rel-f1 dataset). Higher is better. Results of our hybrid model are shown in bold. All baseline scores are taken from the respective publications \cite{wang2025griffin,fey2025kumo}, while the score for the Griffin-pretrained variant was obtained by evaluating the official model checkpoint on our task.}
    \label{tab:results}
\begin{tabular}{llc}
    \toprule
    \textbf{Category} & \textbf{Model} & \textbf{ROC-AUC} \\
    \midrule
    Supervised & LightGBM & 68.86 \\
    & Data Scientist & 69.80 \\
    & RDL & 72.62 \\
    \midrule
    Foundational & KumoRFM (in-context) & 82.41 \\
    & KumoRFM (fine-tuned) & 82.63 \\
    & Griffin (pre-trained) & 59.64 \\
    & Griffin (fine-tuned) & 70.91 \\
    & \textbf{Hybrid (frozen)} & \textbf{61.40} \\
    & \textbf{Hybrid (fine-tuned)} & \textbf{67.40} \\
    \bottomrule
\end{tabular}
\end{table}

Our fine-tuned hybrid model achieves a competitive score of 67.40, approaching the performance of established supervised methods like LightGBM (68.86) and manual feature engineering by data scientists (69.80). These supervised approaches have long been regarded as the gold standard for relational prediction tasks. Furthermore, our fine-tuned model reaches within 5.22 points of the fully supervised, per-database RDL (72.62). This indicates that our hybrid architecture, despite being pre-trained on a limited RelBench subset, effectively bridges the gap to heavily optimized, task-specific baselines.

The comparison with Griffin is particularly revealing regarding the importance of relational context. While our fine-tuned hybrid model (67.40) remains below Griffin's fine-tuned variant (70.91), the most notable result emerges in the frozen-backbone transfer setting. Our frozen hybrid model (61.40) explicitly outperforms the pre-trained Griffin variant (59.64). The Griffin-pretrained variant, by design pre-trained exclusively on single-table data without exposure to relational structure, struggles to adapt to the interconnected structure of relational databases. By explicitly pre-training on REGs, our model circumvents this limitation, underscoring how critical relational structure awareness is for downstream performance.

KumoRFM (82.41 to 82.63) remains significantly ahead of our approach. However, this gap is expected given KumoRFM's advanced graph transformers, multi-modal encoders, and vastly larger training scale. Moreover, it is a closed-source commercial product, making direct comparison under identical conditions impossible. We therefore interpret KumoRFM primarily as an empirical upper bound that demonstrates what is achievable with sufficient scale and dedicated resources.

\subsection{Ablation Study}
\label{subsec:ablation-study}

Table~\ref{tab:ablation} isolates the contribution of each component by varying embeddings (BART vs.\ random) and GNN usage (none, frozen, fine-tuned).

The BART + GNN combination consistently outperforms all other configurations. Without the GNN, BART-only performance drops sharply to an ROC-AUC of 43.90, demonstrating that the MLP head alone cannot compensate for absent relational context. Random embeddings with a fine-tuned GNN reach only 58.00, confirming that BART pre-training provides substantial semantic value beyond what the GNN alone can recover. Fine-tuning the GNN further improves the best configuration from 61.40~to~67.40.

\begin{table}[ht]
\centering
\caption{Ablation results on the test set. "BART" denotes fine-tuned BART embeddings; "Random Emb." denotes embeddings sampled from the same value range as the BART embeddings. "No fine-tuning" keeps GNN weights frozen during downstream training; "with fine-tuning" allows them to be updated. Accuracy and F1 are omitted as they collapse to static values (70.50 and 82.70, respectively) across all configurations due to the majority-class prior ($\approx 70.5\%$). ROC-AUC captures the actual differences in ranking quality.}
\label{tab:ablation}
\begin{tabular}{lcc}
\toprule
\textbf{Model} & \textbf{Prec.} & \textbf{ROC-AUC} \\
\midrule
BART + GNN (no fine-tuning) + Head & 79.60 & \textbf{61.40} \\
BART + GNN (with fine-tuning) + Head & 83.20 & \textbf{67.40} \\
BART + Head & 68.30 & 43.90 \\
Random Emb.\ + Head & 74.20 & 55.40 \\
Random Emb.\ + GNN (no fine-tuning) + Head & 74.90 & 53.50 \\
Random Emb.\ + GNN (with fine-tuning) + Head & 76.00 & 58.00 \\
\bottomrule
\end{tabular}
\end{table}

BART embeddings without the GNN (ROC-AUC 43.90) perform worse than random embeddings without the GNN (55.40). We attribute this to a domain shift effect: the BART encoder was fine-tuned on six other RelBench databases and never exposed to rel-f1. Its embeddings may therefore occupy a narrow region of the feature space that is poorly aligned with the downstream task, making it difficult for the MLP head to extract a linearly separable signal. Random embeddings, by contrast, are uniformly distributed and provide a more neutral starting point for the linear operations of the classifier. Crucially, while BART embeddings preserve relative semantic structure (semantically similar rows lie close to one another), the MLP head relies on linearly separable absolute coordinates and struggles to exploit this latent structure. This relative geometry is exactly the signal that message passing amplifies: once the GNN is introduced, the ordering is reversed (61.40 vs.\ 53.50 for frozen; 67.40 vs.\ 58.00 for fine-tuned), confirming that the GNN can propagate and amplify the latent semantic structure that the MLP head alone cannot recover.

Due to the severe class imbalance (majority-class prior of $\approx 70.5\%$), accuracy and F1 computed at the default threshold collapse to static values (70.50 and 82.70, respectively) across all configurations. We therefore omit them from our reported results. While precision reflects minor differences in the sparse positive-class predictions, ROC-AUC, being threshold-independent, most reliably captures the actual differences in ranking quality and serves as our primary metric.

\begin{figure}[htbp]
    \centering
    \includegraphics[width=0.9\linewidth]{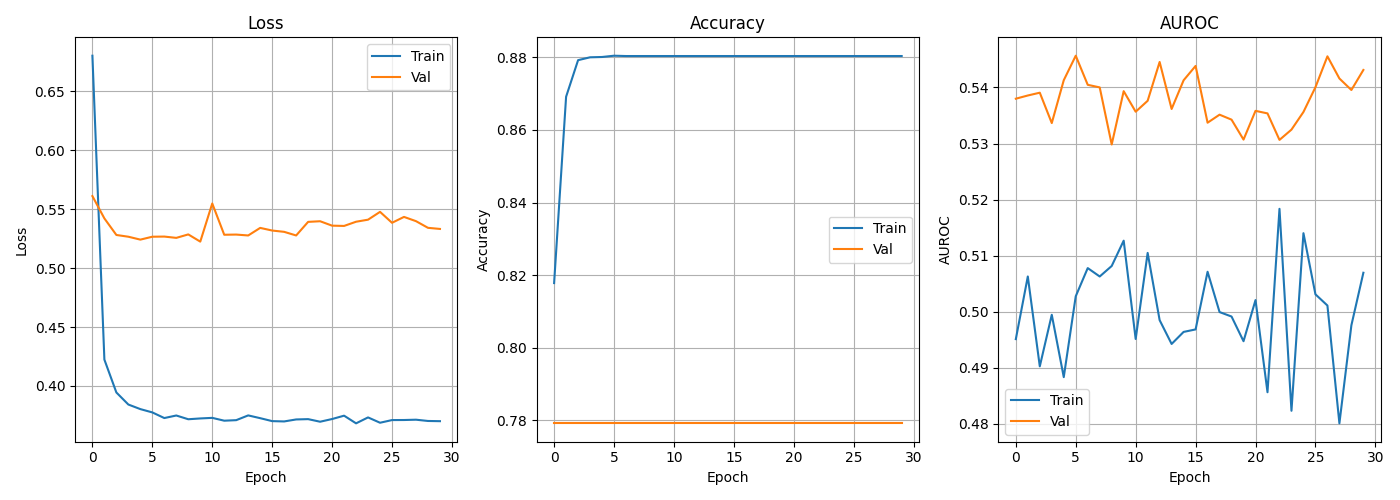}
    \caption{Training and validation metrics (loss, accuracy, and ROC-AUC) on the rel-f1 downstream task with frozen GNN parameters. The validation ROC-AUC exhibits considerable instability and does not show a clear upward trend.}
    \label{fig:loss-f1-no-finetune}
\end{figure}

\begin{figure}[htbp]
    \centering
    \includegraphics[width=0.9\linewidth]{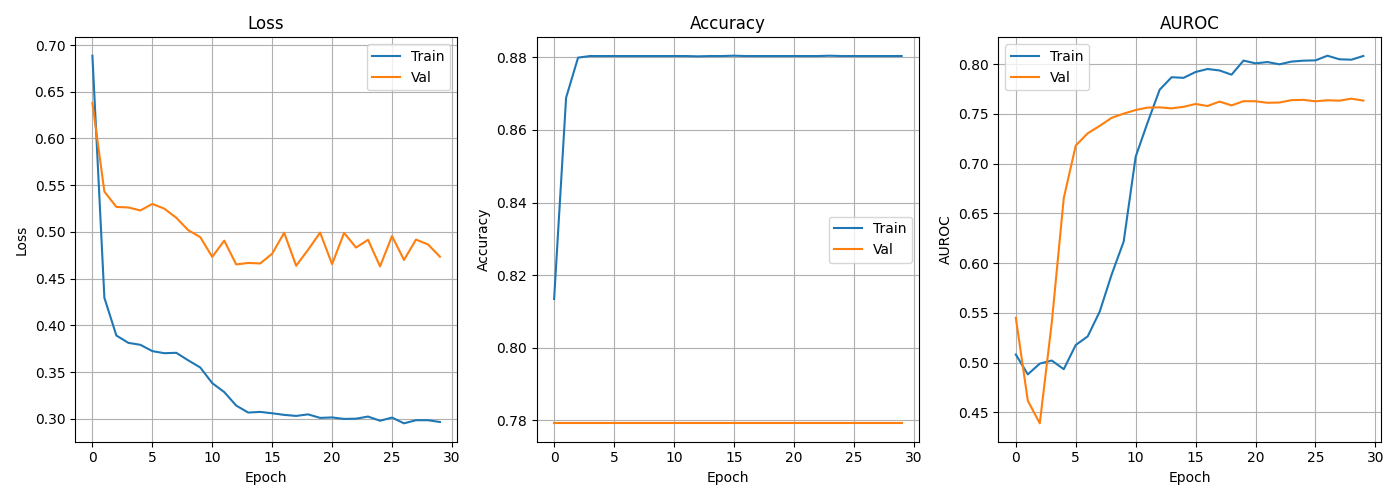}
    \caption{Training and validation metrics (loss, accuracy, and ROC-AUC) on the rel-f1 downstream task with learnable GNN parameters. In contrast to the frozen setting (Fig.~\ref{fig:loss-f1-no-finetune}), validation ROC-AUC improves steadily across epochs.}
    \label{fig:loss-f1-finetune}
\end{figure}

The training dynamics (Figs.~\ref{fig:loss-f1-no-finetune} and~\ref{fig:loss-f1-finetune}) corroborate these findings. With GNN fine-tuning, ROC-AUC improves steadily across epochs, indicating that the model progressively adapts its relational representations to the new database. In contrast, the frozen setup produces unstable validation curves without a clear upward trend, suggesting that the pre-trained GNN representations require further adaptation to fully benefit the downstream task.

In summary, these results demonstrate that (i)~BART pre-training provides strong row-level semantic features, (ii)~the GNN effectively enriches them with relational context through message passing, and (iii)~fine-tuning the pre-trained GNN is particularly beneficial when adapting to unseen relational databases.

\section{Discussion}
\label{sec:discussion}

Building upon our experimental findings, we first analyze the roles of BART as a row encoder and the GNN as a relational encoder, examining how they complement each other. We then address the current limitations of our framework and identify the gaps that must be bridged before it can serve as a true relational foundation model.

\begin{figure}[htbp]
    \centering
    \includegraphics[width=0.8\linewidth]{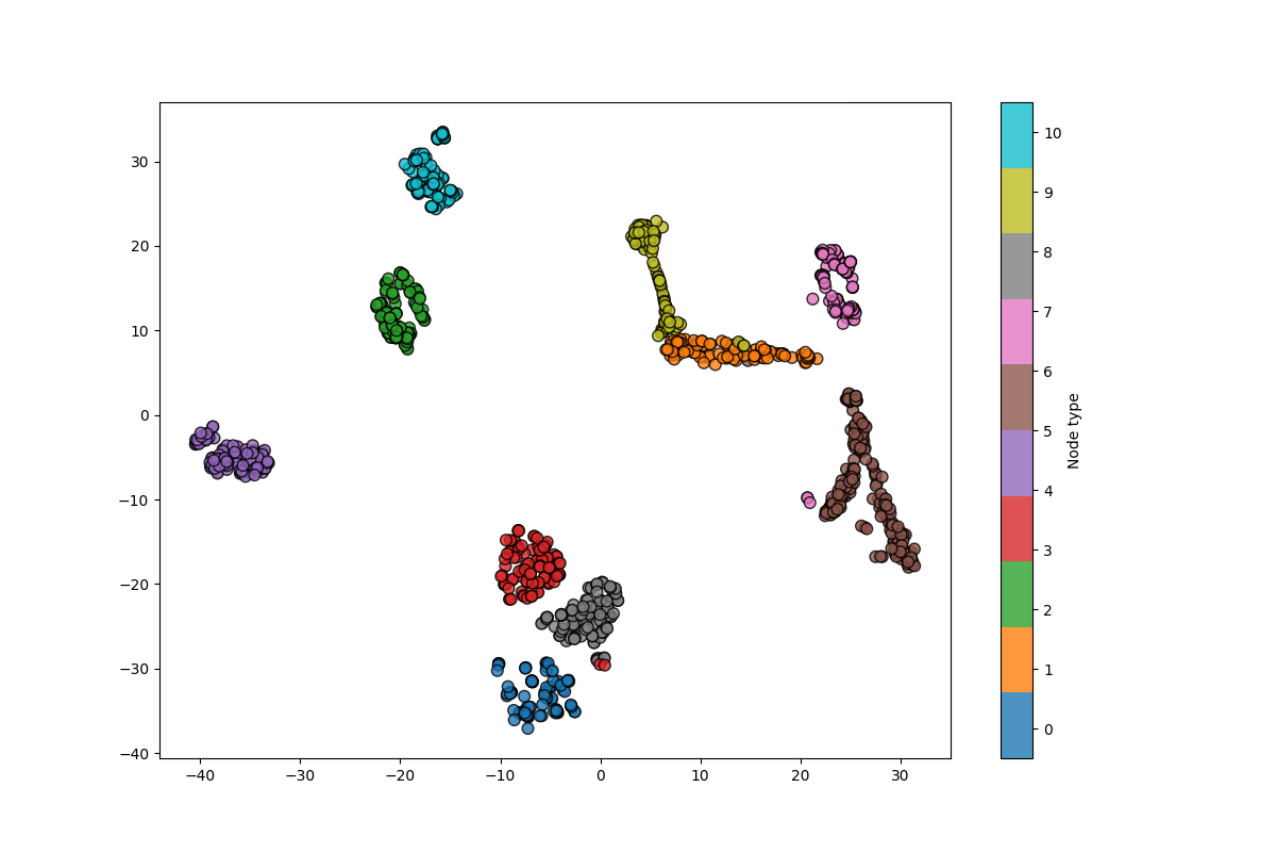}
    \caption{A t-SNE visualization of row embeddings after BART encoding (before GNN processing) for the \textit{rel-f1} database. Each color corresponds to a distinct node type within the generated heterogeneous graph (e.g., \texttt{drivers}, \texttt{races}). The encoder produces coherent type-specific clusters, although some expected overlap is visible.}
    \label{fig:t-sne-bart}
\end{figure}

\paragraph{BART as Row Encoder.}
Using BART to encode relational rows leverages linguistic pre-training to capture attribute-value dependencies and offers a streamlined and unified alternative to type-specific multi-modal encoders. Fig.~\ref{fig:t-sne-bart}, produced with t-SNE \cite{vandermaaten2008tsne}, confirms that the fine-tuned encoder groups same-table rows into coherent clusters, though with some overlap and outliers. However, BART alone is limited to intra-row semantics (ROC-AUC 43.90, Table~\ref{tab:ablation}). It inherently lacks the mechanism to explicitly model cross-table dependencies and faces strict sequence length constraints (e.g., a maximum of 1024 tokens), which restricts its applicability to extremely wide tables. These limitations strongly motivate the subsequent GNN stage.

\begin{figure}[htbp]
    \centering
    \includegraphics[width=0.8\linewidth]{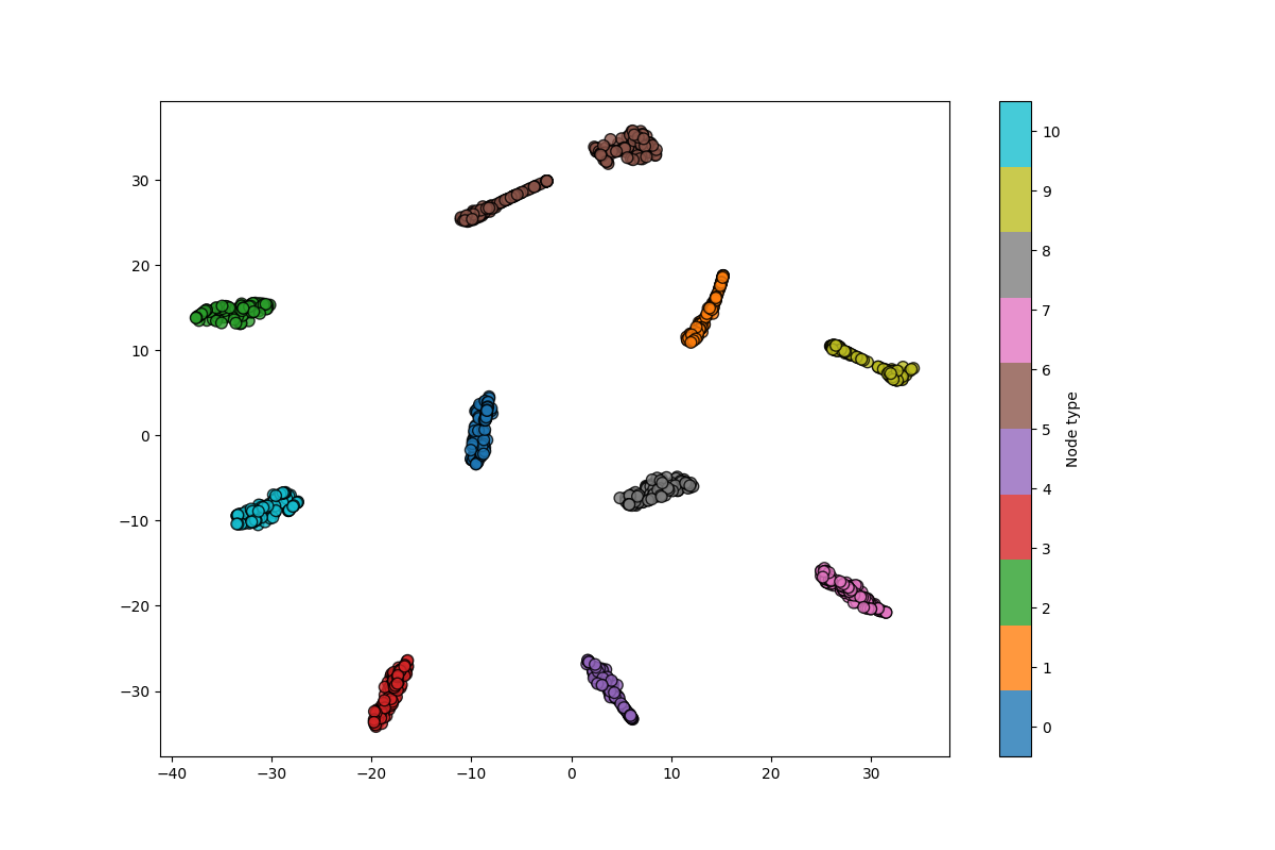}
    \caption{A t-SNE visualization of row embeddings after GNN message passing for the \textit{rel-f1} database. The color coding matches that of Fig.~\ref{fig:t-sne-bart}. Compared to the initial BART embeddings, the GNN propagation yields tighter and more distinctly separated clusters for each node type, visually demonstrating that the injection of relational context produces highly structured representations.}
    \label{fig:t-sne-gnn}
\end{figure}

\paragraph{GNN as Relational Encoder.}
The GNN compensates by injecting relational context through message passing over REGs. As shown in Fig.~\ref{fig:t-sne-gnn}, after GNN propagation the embeddings form tighter, well-separated per-table clusters with minimal noise, demonstrating that relational context produces more structured representations. The ablation confirms this contribution. Adding the GNN raises ROC-AUC from 43.90 to 61.40 (frozen) or 67.40 (fine-tuned). Trade-offs include computational overhead from message passing over million-node graphs, the risk of oversmoothing with mean aggregation, and the lack of joint LM-GNN training leaving potential cross-modal synergies under-explored.

\paragraph{Limitations.}
Several factors currently prevent this framework from acting as a true relational foundation model. First, the pre-training scale is limited. We used only 100,000 rows for BART and 6 RelBench databases for the GNN, which is substantially smaller than the corpora typically used to train modern foundation models. Second, evaluation is restricted to a single downstream task. Broader assessment across regression, link prediction, and multiple held-out databases is needed to substantiate generalization claims. Third, our deliberately simple architecture (single LM encoder, GraphSAGE with mean aggregation) trades expressiveness for architectural parsimony compared to Griffin's multi-modal encoders \cite{wang2025griffin} or KumoRFM's graph transformers \cite{fey2025kumo}. Fourth, the sequential two-stage training disrupts the gradient flow between the BART encoder and the GNN, preventing the framework from learning a joint distribution across semantic and relational modalities. Fifth, all reported results are based on single experimental runs. While our findings are competitive, establishing statistical significance through multiple trials is paramount~\cite{DBLP:conf/emnlp/ZhuPNXHKS24,DBLP:conf/emnlp/ZhuWWZCKS25} within such densely populated and competitive benchmarks.

\section{Conclusion and Open Questions}

We proposed a lightweight hybrid LM-GNN framework combining a fine-tuned BART encoder for intra-row semantics with a GraphSAGE-based GNN for relational context enrichment. On the \texttt{driver-dnf} task from RelBench's \textit{rel-f1} dataset, the model achieved an ROC-AUC of 67.40. This performance comes within 1.46 to 2.40 points of supervised baselines (LightGBM, Data Scientist) and is 5.22 points below RDL. Notably, we achieved these competitive results using a modest budget of 100,000 rows for BART fine-tuning and 6 REGs totaling 28.7 million nodes and 40.1 million edges for GNN pre-training. These results extend the vision of Vogel et al.~\cite{vogel2023towards} beyond data engineering tasks, demonstrating that hybrid architectures can serve as a blueprint for general-purpose relational models.

Several open questions remain for advancing this framework: (1) Can joint LM-GNN training, for example through end-to-end backpropagation, align semantic and relational representations more effectively than the current sequential pipeline? (2) How does scaling to substantially larger and more diverse pre-training corpora affect downstream performance? (3) Does the architecture generalize to broader task types such as regression and link prediction, and to databases with more complex schemas? While our results already demonstrate the viability of the hybrid LM-GNN paradigm, addressing these questions is critical to transform the current framework from a proof-of-concept into a robust relational foundation model.

\begin{acknowledgments}
The authors thank the International Max Planck Research School for Intelligent Systems (IMPRS-IS) for supporting Jingcheng Wu. Jingcheng Wu and Ratan Bahadur Thapa have been funded by the Deutsche Forschungsgemeinschaft (DFG, German Research Foundation) - SFB 1574 - Project number 471687386.
\end{acknowledgments}

\section*{Declaration on Generative AI}
During the preparation of this work, the authors utilized Gemini for the purpose of a grammar and spelling check. The use of Generative AI was strictly confined to minor linguistic refinements to improve the clarity of the original text.

\bibliography{main}

\end{document}